\documentclass[pra,preprint,showpacs,showkeys,amsfonts]{revtex4}
\usepackage{graphicx}
\RequirePackage{times}
\RequirePackage{courier}
\RequirePackage{mathptm}
\begin{document}

%\def\frak{\cal }
%\sloppy

\title{Testing the bounds on quantum probabilities}
\author{Stefan Filipp}
\email{sfilipp@ati.ac.at}
\affiliation{Atominstitut der {\"{O}}sterreichischen Universit{\"{a}}ten,
Stadionallee 2, A-1020 Vienna, Austria}
\author{Karl Svozil}
\email{svozil@tuwien.ac.at}
\homepage{http://tph.tuwien.ac.at/~svozil}
\affiliation{Institut f\"ur Theoretische Physik, University of Technology Vienna,
Wiedner Hauptstra\ss e 8-10/136, A-1040 Vienna, Austria}

\begin{abstract}
Bounds on quantum probabilities and expectation values are derived for experimental setups
associated with Bell-type inequalities.
In analogy to the classical bounds, the quantum limits are experimentally testable
and therefore serve as criteria for the validity of quantum mechanics.
\end{abstract}


\maketitle

\section{Introduction}

Suppose someone claims
that the chances of rain in Vienna and Budapest are 0.1
in each one of the cities alone, and
the joint probability of rainfall in both cities is 0.99.
Would such a proposition appear reasonable?
Certainly not, for even intuitively it does not make much  sense to claim
that it rains almost never in one of the cities, yet almost always in both of them.
The worrying question remains:
which numbers could be considered reasonable and consistent?
Surely, the joint probability should not exceed any single probability.
This certainly appears to be a necessary condition, but is it a sufficient one?
In the middle of the 19th century George Boole,
in response to such queries, formulated a theory of
``conditions of possible experience'' \cite{Boole,Boole-62}
which dealt with this problem.
Boole's requirements on the (joint) probabilities of logically connected events
are expressed by certain equations or inequalities relating those (joint) probabilities.

Since Bell's investigations \cite{bell-87,clauser,peres} into bounds on classical probabilities,
similar inequalities for a particular physical setup have
been discussed in great number and detail.
In what follows, the classical bounds are referred to as ``Bell-type inequalities.''
Whereas these bounds are interesting
if one wants to inspect
the violations of classical probabilities by quantum probabilities,
the validity of quantum probabilities and their experimental verification is a completely
different issue.
Here we shall present detailed numerical studies on the bounds of quantum probabilities which,
in analogy to the classical bounds, are experimentally testable.

\subsection{Correlation Polytopes}

In order to establish bounds on quantum probabilities, let us recall that
Pitowsky has given a geometrical interpretation of the bounds of classical probabilities
in terms of correlation polytopes \cite{pitowsky-86,pitowsky,pitowsky-89a,Pit-91,Pit-94}
[see also Froissart \cite{froissart-81} and
Tsirelson (also spelled Cirel'son) \cite{cirelson:80,cirelson}].

Consider an arbitrary number of classical events $a_1, a_2,\ldots , a_n$.
Take some (or all of) their probabilities
and some (or all of) the joint probabilities
$p_1, p_2,\ldots , p_n, p_{12},\ldots $
and identify them with the components of
a vector  $p=(p_1, p_2,\ldots , p_n, p_{12},\ldots )$
formed in Euclidean space.
Since the probabilities $p_i$, $i=1,\ldots ,n$ are assumed to be independent,
every single one of their extreme cases $0,1$ is feasible.
The combined values of $p_1, p_2,\ldots , p_n$ of the extreme cases $p_i=0,1$,
together with the joined probabilities $p_{ij} =p_i p_j$
can also be interpreted as rows of a truth table; with $0,1$ corresponding to
``{\it false}'' and
``{\it true,}'' respectively.
Moreover, any such entry corresponds to a {\em two-valued measure}
(also called {\em valuation, 0-1-measure} or {\em dispersionless measure}).

In geometrical terms,
any classical probability distribution is representable by some convex sum over
all two-valued measures characterized by the row entries of the truth tables.
That is, it corresponds to
some point on the face of the classical correlation polytope $C={\rm conv} (K)$
which is defined by the set of all points whose
convex sum extends over all vectors associated with row entries in the truth table $K$.
More precisely,
consider the convex hull
${\rm conv} (K)=\left\{ \sum_{i=1}^{2^n} \lambda_i{\bf x}_i
  \; \left|  \;
\lambda_i\ge 0,\; \sum_{i=1}^{2^n}\lambda_i =1
\right.
\right\} $
of the set
$$K
=\{{\bf x}_1,{\bf x}_2,\ldots ,{\bf x}_{2^n}\}
= \left\{
\left.
\large(t_1, t_2,\ldots , t_n, t_xt_y,\ldots \large)
\; \right| \;
t_i \in \{0,1\},\; i=1,\ldots ,n
\right\}.$$
Here, the terms $t_xt_y,\ldots$ stand for arbitrary products associated with
the joint propositions which are considered. Exactly what terms  are
considered depends on the particular physical configuration.

By the Minkoswki-Weyl representation theorem \cite[p.29]{ziegler},
every convex polytope has a dual (equivalent) description:
(i)
either as the convex hull of its extreme points; i.e., vertices;
(ii)
or as the intersection of a finite number of half-spaces,
each one given by a linear inequality.
The linear inequalities,
which are obtained
from the set $K$ of vertices
by solving the so called {\em hull problem}
coincide with Boole's ``conditions of possible experience.''

For particular physical setups,
the inequalities can be identified with Bell-type inequalities which have to be satisfied by
all classical probability distributions.
These conditions are demarcation criteria; i.e.,
they are complete and maximal in the sense that no other system of inequality exist
which characterizes the correlation polytopes completely and exhaustively
(That is, the bounds on probabilities cannot be enlarged and improved).
Generalizations to the joint distributions of more than two particles are straightforward.
Correlation polytopes have provided a systematic, constructive way of finding
the entire set of Bell-type inequalities associated with any particular physical configuration
\cite{2000-poly,2001-cddif}, although from a computational complexity point of view
\cite{garey}, the problem remains intractable \cite{Pit-91}.

\subsection{Quantum Probabilities}

Just as the Bell-type inequalities represent bounds on the
classical probabilities or expectation values,
there exist bounds on quantum probabilities.
In what follows we shall concentrate on these
quantum plausibility criteria, in particular on the bounds
characterizing the demarcation line for quantum probabilities.

Although being less restrictive than the classical probabilities,
quantum probabilities do not violate the
Bell-type inequalities maximally
\cite{pop-rohr,mermin-1995,svozil-krenn}.
Tsirelson \cite{cirelson:80,cirelson,khalfin-97}
as well as Pitowsky \cite{pit:range-2001}
have investigated the analytic aspect of bounds on quantum
correlations.
Analytic bounds can also be obtained {\it via} the
{\em minmax principle} \cite[\S 90]{halmos-vs}, stating that
the the bound (or norm) of a self-adjoint operator is equal to the maximum
of the absolute values of its eigenvalues.
The eigenvectors correspond to pure states associated with these eigenvalues.
Thus, the minmax principle is for the quantum correlation functions
what the Minkoswki-Weyl representation theorem is for the classical correlations.
Werner and Wolf \cite{werner-wolf-2001}, as well as Cabello \cite{cabello-2003a},
have considered maximal violations
of correlation inequalities, and have also enumerated quantum states
associated with extreme points of the convex set of quantum correlation functions.

% M. Reed and B. Simon, Methods of Modern Mathematical Physics IV: Analysis of Operators (Academic Press, New York, 1978) Sections XIII.1 and XIII.2.

The maximal violation of the
Clauser-Horne-Shimony-Holt (CHSH) inequality involving expectation values of
binary observables is related to Grothendieck's constant
\cite{fishburn-reeds-1994}. But the demarcation criteria for quantum
probabilities are still far less understood than their classical
counterparts.
In a broader context, Cabello has described a
violation of the CHSH inequality beyond the quantum mechanical (Tsirelson's)
bound by applying selection schemes to particles in a GHZ-state
\cite{cabello-02a,cabello-02b}, yet here we only deal with the usual quantum probabilities of events
which are not subject to selection procedures.

To be more precise,
consider the set of all  single particle
probabilities
$q_{i}={\rm tr}[W(E_{i}\otimes {\Bbb I})]$
and
${\rm tr}[W({\Bbb I}\otimes F_{i})]$,
as well as the two particle joint probabilities
$q_{ij}={\rm tr}[W(E_{i}\otimes F_{j})]$, where some $E_{i}$,
$F_{j}$ are projection operators on a Hilbert space $H$,
and $W$ is some state on $H\otimes H$.
Again, generalizations to the joint distributions of more than two particles are straightforward.
An analogue to the classical correlation polytope  $C$
is the set of all quantum probabilities
\begin{equation}
\label{e-2003-qpoly-Q}
\begin{array}{l}
Q
= \left\{
\left.
\large(q_1, q_2,\ldots , q_n, q_{xy},\ldots \large)
\; \right| \;
q_{i}={\rm tr}[W(E_{i}\otimes {\Bbb I})]\;{\rm or}\;\;{\rm tr}[W({\Bbb I}\otimes F_{i})],\; q_{ij}={\rm tr}[W(E_{i}\otimes F_{j})]
\right\},
\end{array}\nonumber
\end{equation}
with $E_{i}E_{i}=E_{i},\; F_{j}F_{j}=F_{j}$,
$W^\dagger =W,\; {\rm tr} (W)=1$,
and
$\langle u | W | u \rangle \ge 0$,
$i,j=1,\ldots ,n$ for all $\vert u\rangle \in H\otimes H$.
The vertices of classical correlation polytopes $C$
coincide with points of $Q$,
if $E_i,F_j\in \{{\rm diag}(0,\ldots ,0),{\Bbb I}={\rm diag}(1,\ldots ,1)\}$, where
${\rm diag}(a,b,\ldots )$ stands for the diagonal matrix
with diagonal entries $a,b,\ldots $;
in these cases, $W$ may be arbitrary.
A proof of the convexity of $Q$ can be found in  \cite{pit:range-2001}.
Notice, however, that geometrical objects derived from expectation values
need not be, and in fact are not convex, as an example below shows.

One could obtain an intuitive picture of $Q$ by imagining
it as an object (in high dimensions) created from ``soap surfaces''
which is suspended on the edges of
$C$, and which is blown up with air: the original polytope
faces which are hyperplanes get ``bulged''
or ``curved out'' such that,
instead of a single plane
per face,
a continuity of tangent hyperplanes are
necessary to characterize it
\cite{khalfin-97}.

\section{Numerical studies}

In what follows, we shall first consider
the parameterization of projections and states.
The numerically calculated expectation values obey
the Tsirelson bound, exceeding the
values for the classical Clauser-Horne-Shimony-Holt (CHSH) inequality.
Then, we shall deal with the Clauser-Horne (CH) inequality and a higher
dimensional example taken from \cite{2000-poly} in more detail,
followed by an attempt to depict the convex body $Q$ itself.

\subsection{Parameterization}

\begin{figure}%[!htbp]
  \centering
\includegraphics[width=90mm]{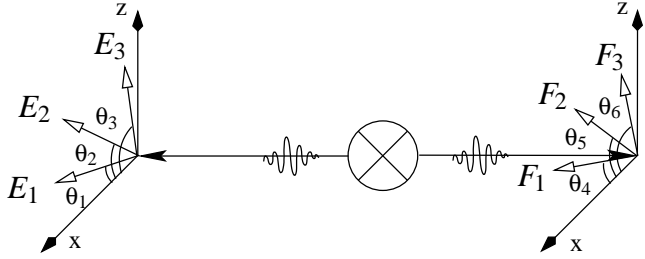}
  \caption{Measurements of spin components corresponding to the projections  ${E}_i$ and ${F}_j$.}
  \label{f-2003-qpoly-1}
\end{figure}

Consider a two spin-1/2 particle configuration,
in which the two particles move in opposite directions along the $y$-axis,
and the spin components are measured in the $x$--$z$ plane,
as depicted in Figure \ref{f-2003-qpoly-1}.
In such a case, the single-particle
 spin observables along $\theta$ correspond to the projections
$E_{i}$ and $F_{j}$; i.e., $E_i,F_j=E(\theta_i),F(\theta_j)$ with
\begin{equation}
  \label{e-2003-qpoly-1}
E(\theta )=F(\theta )=\frac{1}{2}({\Bbb I} +{\bf n}(\theta)\cdot{\bf \sigma})=
\frac{1}{2}\left(
  \begin{array}{cc}
    1 + \cos\theta& \sin\theta\\
    \sin\theta&1 - \cos\theta
    \end{array}
\right),
\end{equation}
where ${\bf \sigma}$ is the vector composed from the Pauli spin matrices.

Any state represented by the operator $W$ must be
(i) self-adjoint $W^\dagger =W$,
(ii) of trace class ${\rm tr} (W)=1$, and
(iii) positive semidefinite
$\langle u | W | u \rangle \ge 0$
(in another notation, $u^\dagger Wu\ge 0$)
for all vectors $u\in H\otimes H$.
For the state to be pure, it must be a projector $W^2=W$,
or equivalently, ${\rm tr}(W^2)=1$.

In order to be able to parameterize $W$,
we recall (e.g., \cite[\S 72]{halmos-vs})
that a necessary and sufficient condition for
positiveness is the representation as the square of
some self-adjoint $B$; i.e., $W=B^2$.
In $n$ dimensions, $B$ can be parameterized by $n^2$
real independent parameters. Finally, $W$ can be normalized
by $W/ {\rm tr}(W)$.
Thus, for a two particle problem associated with $n=4$,
\begin{equation}
    \label{e-2003-qpoly-2}
    W=\frac{1}{\sum_{i=1}^4 b_i^2 + 2 \sum_{j=5}^{16} b_j^2}\left(\begin{array}{cccc}
        b_1 & b_5 + {\rm i}b_6 & b_{11} + {\rm i}b_{12} & b_{15} + {\rm i}b_{16}\\
        b_5 - {\rm i}b_6 & b_2 & b_7 + {\rm i}b_8 & b_{13} + {\rm i}b_{14}\\
        b_{11} - {\rm i}b_{12} & b_7 - {\rm i}b_8 & b_3 & b_9 + {\rm i}b_{10}\\
        b_{15} - {\rm i}b_{16} & b_{13} - {\rm i}b_{14} & b_9 - {\rm i}b_{10} &b_4
      \end{array}\right)^2
\end{equation}
for $b_1,b_2,\ldots ,b_{16}\in {\Bbb R}$.

The probability for finding the left particle in the spin-up state
along the angle $\theta_i$ is given  by
$q_i = {\rm tr}\{{W} [{E}(\theta_i) \otimes {\Bbb I}]\}$.
$q_j = {\rm tr}\{{W} [{\Bbb I} \otimes {F}(\theta_j)]\}$ is
the probability for finding the particle on the right hand side along $\theta_j$ in the
spin up state.
$q_{ij} = {\rm tr}\{{W} [{E}(\theta_i) \otimes {F}(\theta_j)]\}$ denotes the joint probability
for finding the left as well as the right particle in the spin-up state along $\theta_i$ and $\theta_j$,
respectively.
The associated expectation values are given by
$E(\alpha,\beta)={\rm tr}\{{W} [{\bf \sigma}_\alpha  \otimes
{\bf \sigma}_\beta ]\}$,
where $\sigma_\alpha= {\bf n}(\alpha )\cdot {\bf \sigma}$,
and ${\bf n}(\alpha ),{\bf n}(\beta )$
are unit vectors pointing in the directions
of spin measurement $\alpha$ and $\beta$, respectively.

\subsection{Violations of Bell-type inequalities}
We can utilize  the parameterizations of measurement operators
$E_i,\,F_j$ from Eq. (\ref{e-2003-qpoly-1})
and of states $W$ from Eq. (\ref{e-2003-qpoly-2})
to find violations of Bell-type
inequalities.
The general procedure is to choose a particular set of
projection operators and randomly generate arbitrary states $W$.
Having created a certain number of states, another set of
projection operators can be chosen as measurement operators.
A proper parameterization of the two sets representing samples of
measurement operators and states yields the basis
for expressing the maximal violations which reflect the quantum hull.
The choice of projection operators depending continuously on one
parameter corresponds to a smooth variation of the measurement
directions.

Restriction of the different measurement
directions to the $x$--$z$-plane perpendicular to the propagation direction of the particles
(cf. Fig. \ref{f-2003-qpoly-1})
permits a two-dimensional visualization of
the quantum hull.
An extension to more than one parameter associated with other measurement
directions is straightforwardly implementable.
On inspection we find that, despite the shortcomings in the visualization, no new insights
can be gained with respect to the model calculations presented here.
Thus, we adhere to these elementary
configurations of measurements in the $x$--$z$-plane described above.

\subsubsection{CHSH case}
In a first step, we shall concentrate on the expectation values
rather than on probabilities.
Consider the CHSH-operator
$
  O_{CHSH}(\alpha ,\beta ,\gamma ,\delta )=
\sigma_\alpha \sigma_\gamma + \sigma_\beta \sigma_\gamma +
  \sigma_\beta \sigma_\delta - \sigma_\alpha \sigma_\delta
$ giving raise to a sum of expectation values
$
{\rm tr}[W\cdot O_{CHSH}(\alpha ,\beta ,\gamma ,\delta )]= E(\alpha,\gamma) +
  E(\beta,\gamma) + E(\beta,\delta) - E(\alpha,\delta)$.
Here, $\alpha$, $\beta$ and $\gamma$, $\delta$ denote coplanar measurement
directions on the left and right hand side of a physical setup according to
Figure \ref{f-2003-qpoly-1}, with $\alpha = \theta_1$, $\beta=\theta_2$ and
$\gamma=\theta_4$, $\delta=\theta_5$, respectively.

\begin{figure}[!ht]
  \centering
  \includegraphics{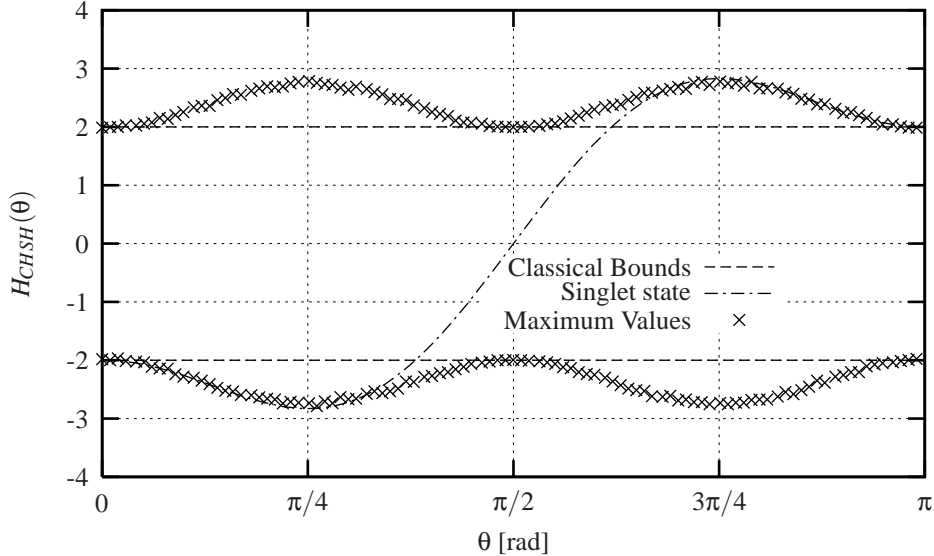}
  \caption{The quantum hull $H_{CHSH}$
as a function of a single parameter $\theta$.}
  \label{f-2003-qpoly-2}
\end{figure}

The quantum expectation values
obey the Tsirelson bound
\footnote{Squaring $O_{CHSH}$ yields  \cite[p. 174]{peres}
$
  O_{CHSH}^2=4+[\sigma_\alpha,\sigma_\beta][\sigma_\gamma,\sigma_\delta]
$.
Since  for any two bounded
operators $A$ and $B$
$
  \|[A,B]\| \leq \|A B\| + \|B A\| \leq 2\|A\|\|B\|
$,
we obtain $\| O_{CHSH}^2 \| \leq 8$ and hence
$  \|O_{CHSH}\| \leq 2\sqrt{2}$.
}
$ \|O_{CHSH}(\alpha ,\beta ,\gamma ,\delta ) \|  \leq 2\sqrt{2}$
for the configuration
$\alpha=0$,
$\beta=2\theta$,
$\gamma=\theta$,
$\delta=3\theta$ along $0 \le \theta \le \pi$.
(The classical CHSH-bound from above is $2$.)
The
particular parameterization include the well-known
measurement directions for obtaining a maximal violation for the
singlet state at $\theta=\pi/4$ and $3\pi/4$.
An analytic expression of the quantum hull for the full
range of $\theta$ is obtained by solving the minmax
problem \cite[\S 90]{halmos-vs} for the CHSH operator; i.e.,
\begin{equation}
H_{CHSH}(\theta )=\pm \sqrt{2[3-\cos (4\theta )]} \le 2\sqrt{2}
.
\end{equation}
The quantum hull $H_{CHSH}$, along
with the singlet state curve, is depicted in Figure \ref{f-2003-qpoly-2}.

\subsubsection{CH case}
Next we  study the quantum hull corresponding to the CH inequality
$ -1\le P_{CH} \leq 0$,
with $
P_{CH} = p_{13} + p_{14} + p_{24} - p_{23} - p_{1} - p_{4}
$.
As this inequality is essentially equivalent to the CHSH inequality
discussed above if the expectation values are expressed by probabilities
\footnote{One can  verify this by assuming physical locality \cite{mermin-1995}
and by inserting the
  relation $E(\alpha,\beta) = 4p_{\alpha\beta} - 1$ into the CHSH
  inequality (see also Cereceda \cite{cereceda-2001}).}, we could in principle produce the same plot as in Figure
\ref{f-2003-qpoly-2} by the same choice of parameterization and a
relabeling of the axes.

Again, the minmax principle yields the analytic expression for the hull; i.e.,
\begin{equation}
H_{CH}(\theta )={1\over 2}\left[\pm {\sqrt{3-\cos (2\theta )\over 2}} - 1 \right]
.
\end{equation}
Thus, in terms of
probabilities,
the upper bound admitted by quantum mechanics is
$H_{CH}(\theta ) \le (\sqrt{2}-1) /2$, corresponding to the Tsirelson bound of $2\sqrt{2}$
in the CHSH case.

\begin{figure}
  \centering
  \includegraphics[clip]{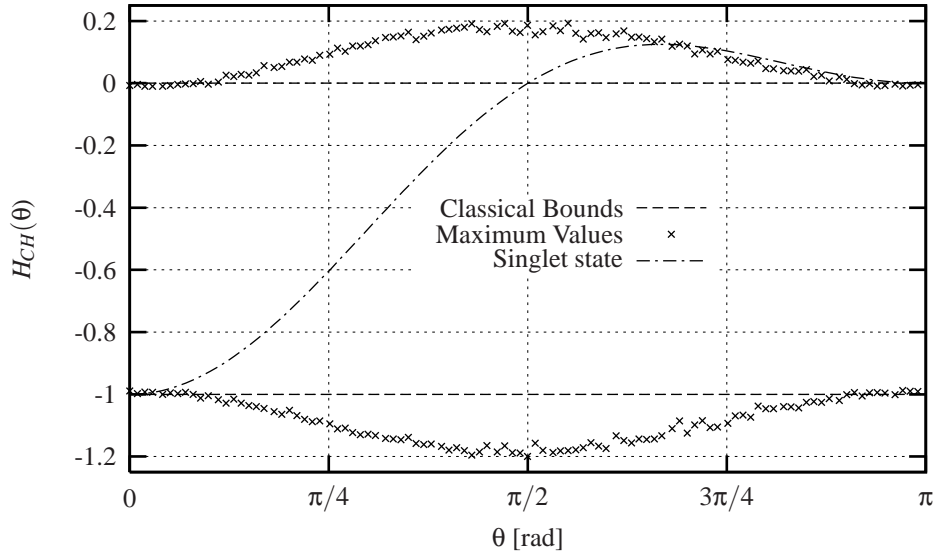}
  \caption{Quantum hull $H_{CH}$
as a function of a single parameter $\theta$.}
  \label{f-2003-qpoly-3}
\end{figure}

To explore the quantum hull also for general configurations where
the singlet state does not violate the inequality maximally,
we restrict the projection operators $E_i,\,F_j$ by
${E}_1(0)$, ${E}_2(\theta)={F}_1(\theta)$, ${F}_2(2\theta)$ to variations of one parameter
$\theta$.
In Figure  \ref{f-2003-qpoly-3} the quantum hull  $H_{CH}$ of $P_{CH}$
obtained by substituting $p$ through $q$ is plotted
along $0 \le \theta \le \pi$. We can observe a maximum at
$\theta=\pi/2$ that does not coincide with the maximum value reached
by the singlet state.

\subsubsection{Two particle three observable case}
\begin{figure}
  \centering
  \includegraphics{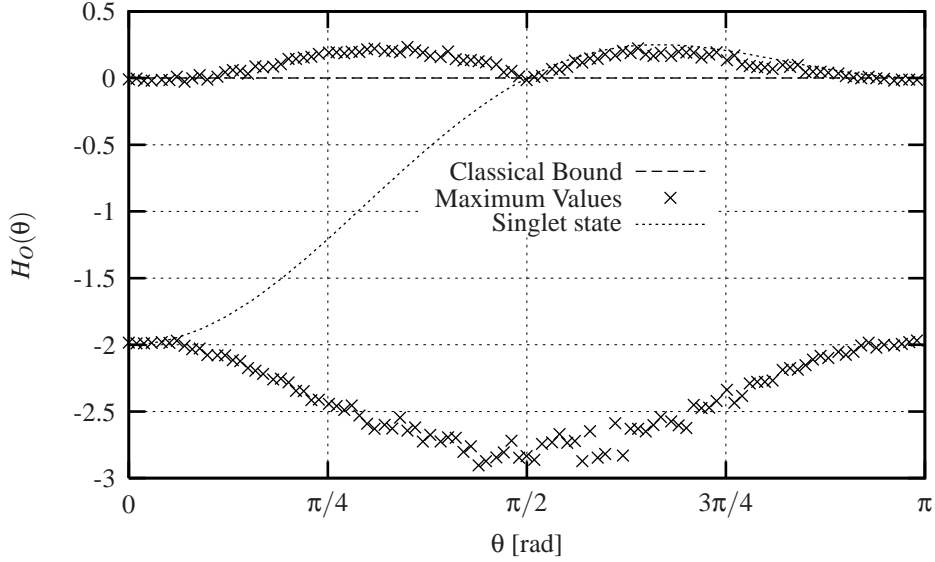}
  \caption{Quantum hull $H_{O}$  as a function of a single parameter $\theta$.}
  \label{f-2003-plotpit}
\end{figure}

As a third example, consider a quantum hull associated with the
configuration involving two spin 1/2 particles and
three measurement directions. One of the 684 Bell-type inequalities enumerated in
\cite{2000-poly} is
$  - p_{14} + p_{15} + p_{16} +
  p_{24} + p_{26} + p_{34} + p_{35} - p_{36} \leq +p_{1}+ p_{2} + p_{4} + p_{5}$.
The associated quantum operator is given by
\begin{equation}
\begin{array}{l}
  O=- {E}_1 \otimes {\Bbb I} - {E}_2 \otimes {\Bbb I} - {\Bbb I} \otimes {F}_1 - {\Bbb I} \otimes
  {F}_2 -
  {E}_1 \otimes {F}_1 + {E}_1 \otimes {F}_2 + {E}_1 \otimes {F}_3 +\\
\qquad \qquad {E}_2
  \otimes {F}_1 + {E}_2 \otimes {F}_3 +
  {E}_3 \otimes {F}_1 + {E}_3 \otimes {F}_2 - {E}_3 \otimes {F}_3.
\end{array}
  \label{e-2003-poly-xyz}
\end{equation}
Taking ${\rm tr}(WO)$ with a symmetric choice of measurement
directions
${E}_1={F}_1=E(0)$,
${E}_2={F}_2=E(\theta)$,
${E}_3={F}_3=E(2\theta)$ ensures a violation of the inequality for the
singlet state at $\theta=2\pi/3$ \cite{2000-poly}.
The associated
quantum hull  $H_{O}$ is depicted in Figure  \ref{f-2003-plotpit}.

The three examples depicted in Figure
\ref{f-2003-qpoly-2}-\ref{f-2003-plotpit} provide tests of the
validity of quantum mechanics in the usual Bell-type inequality setup.
They clearly exhibit a dependence of
the quantum hull on
the measurement directions; i.e.,
 a particular set of
projection operators determines the maximal possible violation of a
Bell-type inequality, although the choice of a state is only
restricted by fundamental quantum mechanical requirements.

\subsection{Quantum Correlation Polytope}

So far, we have considered certain quantum hulls associated with
the faces of classical correlation polytopes, as well as bounds on expectation values,
but we have not yet  depicted
the convex body $Q$ itself.
In what follows, we shall get a view
(albeit, due to the complexity of the contributions to $Q$, a not very sharp one)
of the quantum correlation polytope
for the two particle and two measurement directions per particle configuration.
Note that classically,
the corresponding CH polytope, denoted by $C(2)$, is bound by the $2^4$ vertices
$
(0,0,0,0,0,0,0,0)
$,
$
(0,1,0,0,0,0,0,0),
\ldots
(1,1,1,1,1,1,1,1)
$.
These vertices are also elements of the quantum body $Q(2)$ consisting of vectors
$(q_{1},q_{2},q_3,q_4,q_{13},q_{23},q_{14},q_{24})$ according to
Eq. (\ref{e-2003-qpoly-Q}).

Consider a two-dimensional
cut through the quantum body $Q(2)$ by restricting
$q_{1}=q_{2}=q_{3}=a$ and $q_{13}=q_{14}=q_{24}=b$, $a,b\
\mbox{const.}$; i.~e., by taking vectors of the form
$(a,a,a,q_4,b,q_{23},b,,b)$.
These restrictions allow for a set of states
$W$
and corresponding projection operators $E_i,\,F_j$
\footnote{For this numerical study, also
  the restriction to measurement directions lying in the
  $x$--$z$ plane has been waived, thereby demanding only the separability of the
  projection operators into two subspaces corresponding to the left
  and right hand side of the experimental setup.}
such that six out of
eight quantum probabilities have a definite value and the remaining
probabilities $q_4$ and $q_{23}$ can vary within the quantum
bounds.
Numerically, after generating
arbitrary states and arbitrary projection operators,
a postselection is required for conformity to these restrictions.
To find sufficiently many vectors,
we specify the constants $a,b$ only up to a given tolerance value $\varepsilon$.
More precisely, only states and projection operators yielding
 $q_{1}=q_{2}=q_{3}=a\pm\varepsilon$ and
$q_{13}=q_{14}=q_{24}=b\pm\varepsilon$ for some $a,c$ are chosen.

We have set $a=1/2$, $b=3/8$, and the tolerance to $\varepsilon=\pm
0.015$.
Note that this choice implicates the existence of vectors in $Q(2)$ which are
outside $C(2)$, since the CH-inequality
is violated for $q_{23}<1/8$ and $q_4=1/2$.

\begin{figure}[htbp]
  \centering
  \includegraphics{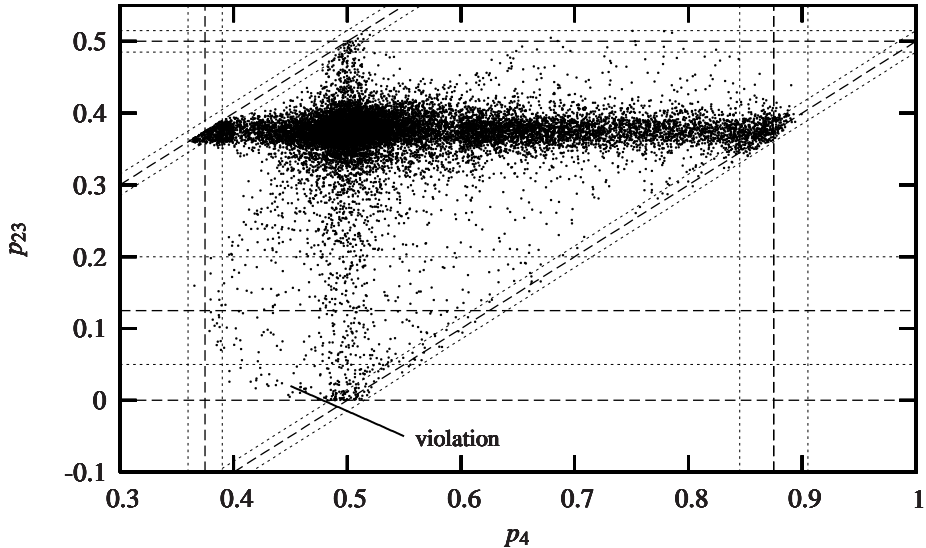}
  \caption{Cut through the quantum body $Q$ for $a=1/2$, $b=3/8$, $\varepsilon=\pm 0.015$.}
  \label{f-2003-bell2hull}
\end{figure}

Figure \ref{f-2003-bell2hull}
depicts a projection of the quantum body $Q(2)$ on
the plane spanned by $q_{4}$ and $q_{23}$. Since the
inequalities constituting the boundary lines have to be modified to
account for $\varepsilon$, the size of $C(2)$ is enlarged
to the dotted lines instead of the dashed lines indicating classical
inequalities. Due to the non-uniform distribution of generated
states, some regions are only sparsely populated.
Nevertheless one can
observe clearly points outside the classical polytope $C(2)$.
We stress the importance of this first glance on $Q(2)$,
since it constitutes the quantum analogy
of the classical correlation polytope $C(2)$,
which has been the
basis of numerous experiments.

\section{Conclusion}

Starting from the correlation polytopes
which represent the restrictions of classical probabilities,
we have used a general parameterization of quantum states and
measurement operators to explore the quantum analogue.
On the basis of the fundamental Bell-type inequalities, the quantum
bounds have been visualized for specific configurations.
We have
presented a
two-dimensional cut through an eight-dimensional quantum body
clearly
exhibiting regions of non-classical probability values.

The quantum bounds predicted in this article
suggest experimental tests in at least two possible forms.
First, our calculations provide an explicit way to construct quantum states, which,
for the measurement setups associated with the orientation of Stern-Gerlach apparatus
or polarizing beam splitters,
yield {\em maximal} violations of the classical bounds by quantized systems.
This is an extension of Tsirelson's original findings \cite{cirelson:80,cirelson}.
Based on the parametrization introduced above, Cabello has proposed such measurements
\cite{cabello-2003a} with a suitable set of maximally entangled states.
These bounds of quantum correlations
have been experimentally tested and verified by Bovino {\it et al.}
\cite{bovino-2003}.

Apart from the concrete experiments mentioned above,
there is a remote possibility of violations of the quantum bounds.
At the moment, these speculations of stronger-than-quantum correlations
\cite{pop-rohr,mermin-1995,svozil-krenn} appear
hypothetical at best, since there is no theoretical indication
that they may be realized physically (besides postselection schemes).
The situation in this respect is clearly different
from the classical bounds in Bell-type inequalities.
Although Bell's inequality does not compare classical probability theory
with a specific theory either, an experimentalist
can utilize these predictions because of
the stronger-than-classical correlations of quantum mechanics.
For instance, in the CHSH case, the experimenter chooses
quantum mechanical setup and preparation procedures such that
the quantum mechanical sum of correlations  violates this bound
most strongly.  Stated pointedly,
Bell's inequality tells the experimentalist what to
measure, but there is no empirical evidence supporting
any experiment to trespass and falsify the quantum bounds.
Nevertheless, it is interesting to know the quantum predictions exactly;
not only from a principal or hypothetical point of view.
Empirical implementations such as the Bovino {\it et al.}
\cite{bovino-2003} experiment test the fine structure of the quantum limits
beyond the  Tsirelson bound.

This research has been supported by the Austrian Science Foundation (FWF), Project Nr. F1513.

%\bibliography{svozil}
%\bibliographystyle{apsrev}
%\bibliographystyle{unsrt}
%\bibliographystyle{plain}

\end{document}